# Tailoring the magnetism of GaMnAs films by ion irradiation

Lin Li[1,2], S D Yao[2], Shengqiang Zhou[1,2*], D Bürger[2], O Roshchupkina[2], S. Akhmadaliev, A W Rushforth[3], R P Campion[3], J Fassbender[2], M Helm[2], B L Gallagher[3], C Timm[4] and H Schmidt[2]

[1]State Key Laboratory of Nuclear Physics and Technology, Peking University, Beijing 100871, China

[2]Institute of Ion Beam Physics and Materials Research, Forschungszentrum Dresden-Rossendorf, P.O. Box 510119, 01314 Dresden, Germany

[3]School of Physics and Astronomy, University of Nottingham, Nottingham NG7 2RD, United Kingdom

[4]Institute for Theoretical Physics, Technische Universität Dresden, 01062 Dresden, Germany

*Email: s.zhou@fzd.de

Abstract

Ion irradiation of semiconductors is a well understood method to tune the carrier concentration in a controlled manner. We show that the ferromagnetism in GaMnAs films, known to be hole-mediated, can be modified by He ion irradiation. The coercivity can be increased by more than three times. The magnetization, Curie temperature and the saturation field along the out-of-plane hard axis all decrease as the fluence increases. The electrical and structural characterization of the irradiated GaMnAs layers indicates that the controlled tailoring of magnetism results from a compensation of holes by generated electrical defects.

**1. Introduction**

For conventional ferromagnetic films, energetic ion irradiation has been demonstrated to be an effective way to tailor the magnetism, e.g. saturation magnetization and magnetic anisotropy, in a controlled manner due to structural modifications [1]. In combination with high-resolution lithography, ion irradiation also provides scalability [2]. Since the discovery of ferromagnetism above 100 K in (III, Mn)–V diluted magnetic semiconductors [3], they have attracted large interest due to their possible applications in spintronics [4, 5]. The ability to locally tune the magnetic properties of magnetic semiconductors is an important issue in future semiconductor devices [6, 7, 8]. The ferromagnetism in GaMnAs is very sensitive to the free hole concentration. As is well known, ion irradiation can be used to render a conducting layer highly resistive (i.e., electrically insulating) through the creation of free-carrier trapping centers [9, 10, 11]. Therefore, the application of ion irradiation to ferromagnetic GaMnAs films would offer an effective way to control the free carrier concentration, and hence the ferromagnetism. However, former works introduced too much defects and the ferromagnetism was completely suppressed [11].

In this work, we have generated hole-compensating defects by means of He-ion irradiation. It turns out that the magnetic parameters, such as coercivity and anisotropy of $Ga_{0.94}Mn_{0.06}As$ films can be modified in a controlled manner without the need to change their Mn concentration [12] or thickness [6]. The well-established ion-irradiation technique provides a possibility for easy lateral patterning as well as for integration with conventional microelectronic devices.

## 2. Experimental

GaMnAs films with a Mn concentration of 6% and a thickness of 25 nm were grown on GaAs (001) substrates by low-temperature molecular beam epitaxy (LT-MBE). He$^+$ ions of 650 keV were used to introduce damage in the epitaxial Ga$_{0.94}$Mn$_{0.06}$As films. Different fluences ranging from $1\times10^{14}$/cm$^2$ to $1\times10^{15}$/cm$^2$ (table 1) were applied to tune the concentration of introduced defects. In order to reduce channeling effects, a tilt angle of 7° between the sample surface normal and the incident beam was employed. Magnetic properties were analyzed using a superconducting quantum interference device (SQUID) magnetometer (Quantum Design MPMS). Magnetotransport measurements with the field applied perpendicular to the sample surface were performed in van der Pauw configuration from 5 to 300 K. High-Resolution X-ray Diffraction (HRXRD) and Rutherford Backscattering/ Channeling (RBS/C) were carried out to clarify the structure of the Ga$_{0.94}$Mn$_{0.06}$As irradiated films.

## 3. Results and discussion

The irradiation leads mostly to the formation of the points-like defects in semiconductors (interstitials, vacancies). The general quantity describing the amount of induced defects is is "displacement per atom (DPA)". In the well-accepted SRIM (Stopping and Range of Ions in Matter) code [13], displacements are the sum of vacancies and replacement collisions. In general application of ion irradiation of semiconductors, vacancies are much more than replacement collisions. Therefore, vacancies are normally used as a quantity to compare the effect of irradiation using different ions (Journal of Applied Physics, Vol. 94, No. 10, pp. 6616–6620, 15 November 2003). Figure 1 reveals the distribution of He$^+$ ions and the calculated vacancies by using SRIM [13]. Nearly all He$^+$ ions go through the GaMnAs layer and lie deep in the GaAs substrate. The irradiation process produces similar quantities of defects in the Ga and As sublattices. The microscopic nature of these defects remains usually beyond of the scope of this paper since the detailed study of the kinetics of defect formation is required. However, in a pervious study it is suggested that the vacancies in the Ga sublattice tend to recombine immediately with the interstitials, since they have opposite charge [14]. There is no previous information on the role of Mn atoms in the formation of complex defects in irradiated GaMnAs. However, previous studies of such defects in non-irradiated samples indicate the possible formation of isolated Mn interstitials with As nearest neighbors and pairs of Mn interstitials with As nearest neighbours [15, 16].

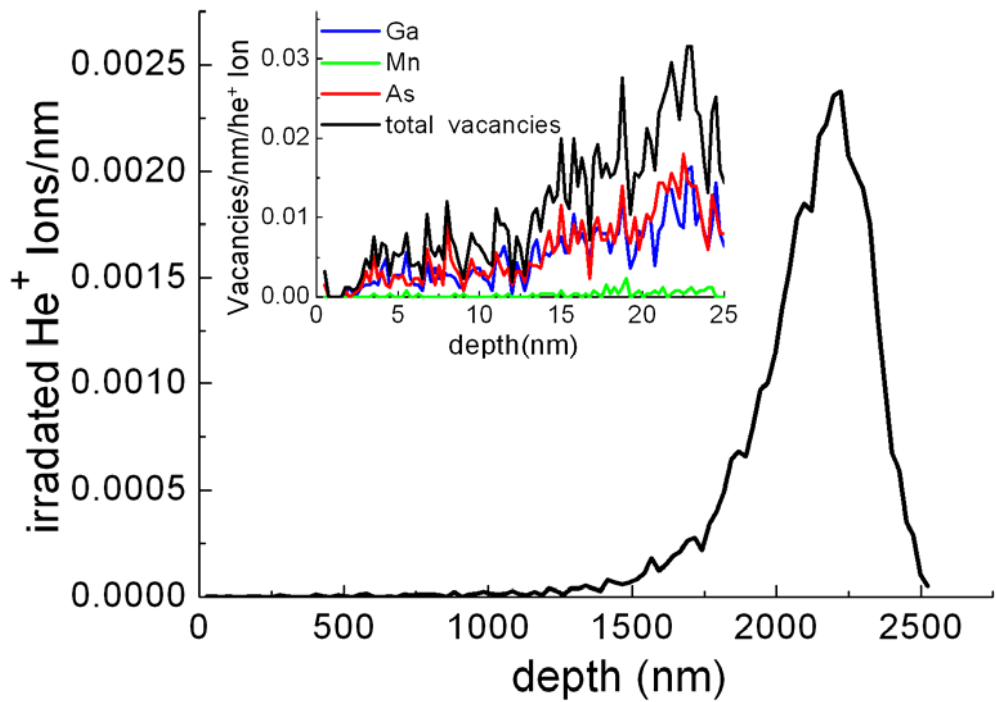

Fig. 1 The distribution of 650 keV He$^+$ ions in 25 nm GaMnAs/GaAs samples. The inset

reveals the calculated number of vacancies by irradiation for Ga, Mn and As elements in the GaMnAs layer.

Figure 2(a) shows the in-plane magnetization at 5 K as a function of the magnetic field. By He-ion irradiation, the coercivity is increased significantly from 50 Oe for the non-irradiated films to 165 Oe when the fluence reaches $3\times10^{14}/cm^2$. At the same time the saturation magnetization decreases slightly from 27 emu/cm$^3$ to 22 emu/cm$^3$. When the fluence increases to $6\times10^{14}/cm^2$ (sample 6E14), the saturation magnetization drops to 13 emu/cm$^3$, and the sample becomes magnetically more isotropic. After applying a He-ion fluence of $1\times10^{15}/cm^2$ (sample 10E14), no ferromagnetism is observed.

Figure 2(b) shows the temperature-dependent remanent magnetization ($M_r$) recorded by first cooling the sample from above the Curie temperature to 5 K under a field of 1 T along $[1\bar{1}0]$ direction, and then setting the field to 1 mT. The remanent magnetization $M_r$ was recorded during warming in a field of 1 mT. For samples 0E14 to 6E14 the $[1\bar{1}0]$ direction represents the easy in plane direction of the magnetization for temperatures close to the Curie temperature. One can see that the Curie temperature drops from 60 K as the He ion fluence is increased.

The free hole concentration (listed in table 1) is estimated by Hall effect measurements at 200 K (not shown here). Although it might be underestimated, due to the influence of the anomalous Hall effect, a clear trend can be observed: The free hole concentration drops as the irradiation fluence increases. This is qualitatively consistent with the increase in resistance (shown later in Figure 3) observed as the irradiation fluence increases. For sample 10E14 (table 1) with the highest He ion fluence, the carrier concentration cannot be determined reliably because of the high resistance, which results from the majority of holes being compensated by irradiation induced defects.

Figure 2(c) shows the out-of-plane normalized magnetization at 5 K for corresponding samples, which reveals that the out-of-plane GaAs [001] direction is the magnetic hard axis, as expected for $Ga_{1-x}Mn_xAs$ grown on (001) GaAs. The observed significant decrease in the saturation field indicates that the out-of-plane hard axis becomes softer with increasing irradiation fluence and finally comparable with the in-plane axis when the fluence is $6\times10^{14}/cm^2$. The shift of the easy axis from in-plane to out-of-plane has also been observed in hydrogenated $Ga_xMn_{1-x}As$ films and was attributed to the drop of free-carrier concentration [17, 18].

Figure 2(d) shows the magnetoresistance (MR=[R(B)-R(0)]/R(0)) with the field applied perpendicular to the samples measured at 5 K. Two factors are considered to contribute to the MR behavior: the positive component at low field below the Curie temperature, which can be attributed to the anisotropic magnetoresistance as the magnetization aligns along the direction of the field [19, 20] and the negative isotropic component at high field, which appears both below and above Curie temperature and has been attributed to the suppression of spin disorder scattering [21, 22, 23]. The negative MR increases largely after irradiation as a result of increased disorder and localization [24, 25]. Strikingly, the negative MR approaches to 100% (the resistivity drops from 3750 $\Omega$cm at 0 T to 25 $\Omega$cm at 6 T) for the sample with the He fluence of $6\times10^{14}/cm^2$ (sample 6E14). The giant MR (100%) is also observed in insulating samples with large Mn concentrations [21, 22]

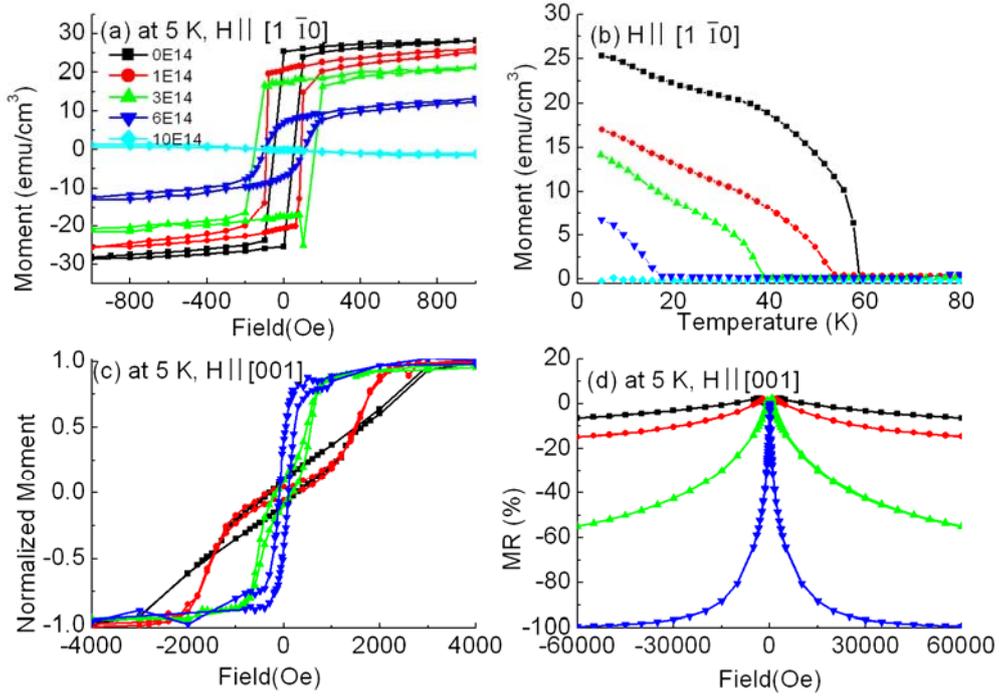

Figure 2. (a) Hysteresis loop at 5 K with the field along the in-plane direction: the coercivity largely increases due to He ion irradiation when the fluence reaches $3\times10^{14}/cm^2$; (b) Temperature dependent remanent magnetization recorded in a field of 1 mT during warming up the samples. (c) Hysteresis loop at 5 K with the field along the out-of-plane direction. The magnetization is normalized for all samples to compare the saturation field, which is decreased with increasing the irradiation fluence. (d) Magnetoresistance (MR) measured with the field perpendicular to the sample surface at 5 K.

Table 1. Implantation fluence, hole concentration and vacancy concentration for all samples

| Samples | 0E14 | 1E14 | 3E14 | 6E14 | 10E14 |
| --- | --- | --- | --- | --- | --- |
| Fluence ($10^{14}/cm^2$) | non-irradiated | 1 | 3 | 6 | 10 |
| Hole concentration($10^{19}/cm^3$) | 3.7 | 3.3 | 1.9 | 0.8 | — |

| Vacancies*($10^{19}$/cm$^3$) | 0 | 1.1 | 3.2 | 6.5 | 11.0 |
|---|---|---|---|---|---|

Figure 3 shows the temperature-dependent resistance of all samples. The as-grown sample and the samples irradiated with low fluences reveal the typical features of metallic conduction observed in ferromagnetic GaMnAs. After irradiation for samples 1E14 and 3E14 the resistance is increased by a factor of three and ten respectively, and for samples 6E14 and 10E14 by several orders of magnitude compared to the non-irradiated sample 0E14. A metal-insulator transition is observed as the fluence increases. In the insert we present the sheet resistance as a function of $T^{-1}$ and $T^{-1/4}$,. None of the samples show signatures of activated transport. Samples with the highest irradiated fluencies show signatures of hopping transport. However, obviously below the Curie temperature it is not possible to extract a reasonable hopping activation energy or thermal activation energy value. The reason could be that the electronic correlations are not included. In the proximity to the metal-insulator transition, electronic correlations may also play a more important role in the underlying mechanism of magnetism of $Ga_{1-x}Mn_xAs$ than previously anticipated [26].

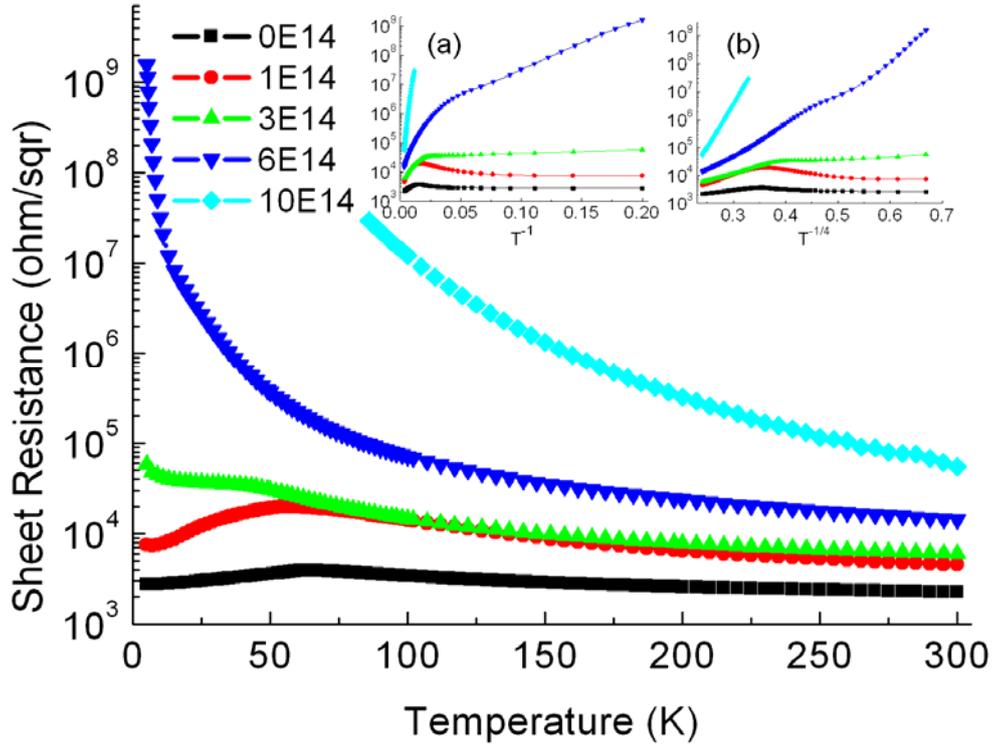

Figure 3. Temperature dependent resistivity of GaMnAs layers after irradiation. The sheet resistance plotted on a logarithmic scale as a function of (inset (a)) $T^{-1}$ and (inset (b)) $T^{-1/4}$.

Figure 4(a) displays the resistivity change at 300 K as a function of the He fluence. Similar phenomena were reported for irradiated p-GaAs, where a small fluence changed the resistivity only slightly, while at higher fluences the resistivity increased rapidly and finally saturates when nearly all free holes were compensated by the generated electron-like defects [10]. The dashed line shows the trend of the resistivity vs. the density of vacancies according to Ref. 10. We estimate that the threshold to reach a saturated resistance for our samples should be more than $1\times10^{15}$ /cm$^2$ [10]. Figure 4(b) shows the compensated hole density versus calculated density of vacancies. In order to compare with former results, we show total vacancies in Figure 4(b). The shaded range roughly separates the non-ferromagnetic vacancy regime from the modified ferromagnetic one. This trend, being independent of ion species and their energy, could be a reference for further work. One can use accessible energetic ions to locally create regions of GaMnAs with different coercivities as well as non-ferromagnetic GaMnAs. Thevenard *et al.* used hydrogen plasma to passivate holes in GaMnAs films. However, the processing cannot be controlled precisely. Indeed, the GaMnAs films were first fully passivated, and only after a second processing step, namely long-time, low-temperature annealing, the magnetic properties of GaMnAs have been modified [12, 17]. Carmeli *et al.* have demonstrated the enhancement of the magnetization of GaMnAs/GaAs superlattices by chemisorption of polar organic molecules, which only allows for controllable states: with and without organic substances [27]. Compared with the attempted approaches, ion irradiation as a one-step processing allows for a more accurate control over the hole concentration in GaMnAs films by several orders of magnitude. Consequently the magnetic properties can be modified in a much more detailed and controlled manner: the anisotropy, the coercivity, Curie temperature and the magnetization can be changed in a *gradual sequence*.

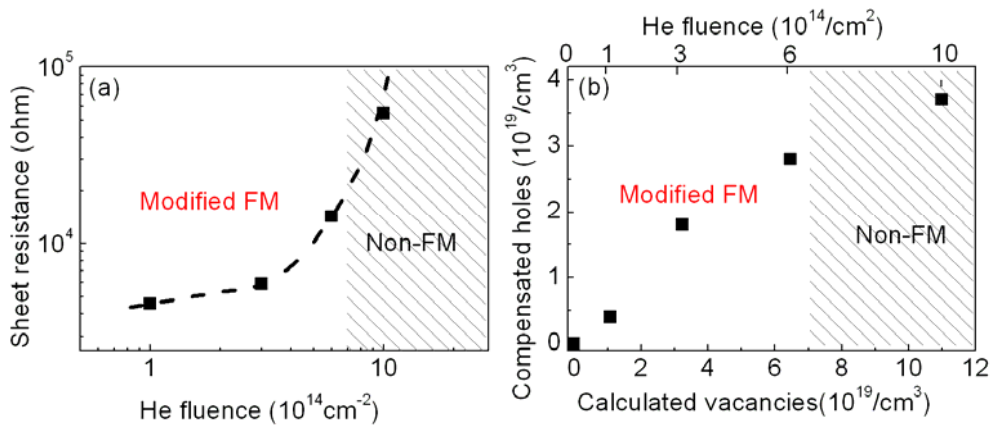

Figure 4. (a) Resistivity at 300 K of samples 1E14-10E14 as a function of the He fluence. The dashed line shows the trend of the resistivity vs. the amount of calculated vacancies according to Ref. 10; (b) compensated holes versus the number of vacancies in the $Ga_{0.94}Mn_{0.06}As$ layer calculated by the SRIM code. The shaded range roughly separates the non-ferromagnetic (Non-FM) vacancy regime from the modified ferromagnetic (FM) one.

HRXRD and RBS/C (not shown) measurements were carried out to check the structural damage. Both measurements indicate that the crystal quality of the epitaxial $Ga_{0.94}Mn_{0.06}As$ films is maintained after irradiation: in the HRXRD spectrum the peak of $Ga_{0.94}Mn_{0.06}As$ layer is almost the same for all samples. From the reciprocal space mapping, the $Ga_{0.94}Mn_{0.06}As$ layer remains fully strained on the GaAs substrate. From RBS/C spectra, the crystal quality remains as good as the virgin one after irradiation. In other words, modifications of the lattice cannot be responsible for the observed effects. Instead, the compensation of holes by generated defects is the dominant origin for the observed change of magnetization due to He ion irradiation.

**4. Conclusions**

In summary, the magnetism of $Ga_{0.94}Mn_{0.06}As$ films can be tailored in a controllable manner by ion irradiation. Defects introduced by energetic ions trap free carriers and modify the magnetic properties. For our study, when the free hole concentration drops to half the value of the as-grown sample, the coercivity is increased more than three times and the saturation field along the out-of-plane hard axis is decreased. This is accompanied by a decrease in the magnetization and the Curie temperature. We demonstrate that ion irradiation provides a way to tailor the magnetism of GaMnAs films without changing the crystalline structure.

## Acknowledgments

L.L. acknowledges financial support by China Scholarship Council, File No.2009601260 for her stay at FZD. S.Z., D.B., and H.S. acknowledge the financial support from the Bundesministerium für Bildung und Forschung (FKZ13N10144). AW. R., RP. C., and BL. G. acknowledge support from EU grant NAMASTE 214499.